# Deep Reinforcement Learning for Multi-Resource Multi-Machine Job Scheduling

Weijia Chen[*†], Yuedong Xu[*], Xiaofeng Wu[*]

[*] Research Center of Smart Networks and Systems, Fudan University, Shanghai, China
[†] School of Operations Research and Information Engineering, Cornell University
wc592@cornell.edu, {ydxu, xiaofengwu}@fudan.edu.cn

*Abstract*—Minimizing job scheduling time is a fundamental issue in data center networks that has been extensively studied in recent years. The incoming jobs require different CPU and memory units, and span different number of time slots. The traditional solution is to design efficient heuristic algorithms with performance guarantee under certain assumptions. In this paper, we improve a recently proposed job scheduling algorithm using deep reinforcement learning and extend it to multiple server clusters. Our study reveals that deep reinforcement learning method has the potential to outperform traditional resource allocation algorithms in a variety of complicated environments.

## I. INTRODUCTION

Resource management is one of the fundamental problems in computer networks and operating systems. The resource allocation problems are usually combinatorial that can be mapped into different NP hard problems. Though each resource allocation scenario can be specific, the general approach is to design efficient heuristic algorithms with performance guarantee under certain conditions. Ghodsi et al. proposed Dominant Resource Fair Queuing (DRF) [1] to assign CPU and memory slots in which one type of resources is deemed as scarce. Wang et al. proposed a Multi-Resource Round Robin (MR$^3$) method [2] that performs packet-level scheduling with an O(1) complexity order. In general, a good resource scheduling algorithm ensures the fairness of coexisting flows as well as the scheduling efficiency.

Beyond the scope of resource management, deep learning method has gained popularity in a large variety of research areas. Google DeepMind combined deep learning with reinforcement learning for artificially intelligent game players and achieved expert-level performance in multiple games. The reinforcement learning agent acts as a decision maker while deep learning method extracts features from state vectors which are in the form of images. An important question arises: *can deep reinforcement learning reshape the resource management in computer networks?* The first successful attempt to our knowledge is DeepRM [3]. In adopting a similar DRL approach as that of Google DeepMind, it represents the resource occupancy status as images so that the deep learning method can work on it, and uses a reinforcement learning agent to make decisions and assign resources to different jobs.

In this paper, we present several improvements to [3] and extend the model with the purpose of identifying the potentials of DRL in multi-resource allocation problems. First, we improved the model by reconstructing the representation of the state space, rewriting the reward function for the reinforcement learning agent, and using a convolutional input layer. We see significant improvement in the learning curve after adopting these approaches. Second, we extend our model to work in a multi-resource multi-machine environment. This is achieved by adding channels in the output layer of the network to represent resources in different machines. The experimental results validate the advantage of using more advanced neural network models and the feasibility of scheduling jobs on multiple machines. These observations will pave the way for our future study on DRL-based resource scheduling in computer networks.

## II. MODEL DESIGN

Consider a cluster with $m$ machines, each having $d$ resource types, and a queue of jobs, each requesting a fixed number of different resources for a fixed time period (e.g. 5 units of CPU and 3 units of memory for 3 timesteps.). At each discrete timestep, incoming jobs arrive and wait in a fixed-length queue. The number of further jobs are saved in a backlog when the queue is full. Meanwhile, the scheduler picks jobs and assigns them to machines. If the machine is busy, the scheduler allocates a job for a machine to process in future time. As time goes, scheduled jobs are processed and the scheduler allocates new jobs as long as the queue is not empty.

**Objective.** Multiple objectives are used for our model. The primary objective, average job slowdown, is defined as $S_i = C_i / T_i$, where $C_i$ is the completion time of the job, and $T_i$ the (ideal) duration of the job. Note that $S_i \geq 1$.

**State Representation.** We represent the state of the system — both the current status of machine resources in the cluster and the resource profiles requested by jobs in the queue — as binary matrices (See Fig. 1. We use colors only to illustrate different jobs. Colored squares represent 1. White squares represent 0) The cluster matrices (two leftmost columns) show the status of scheduled jobs for different machines. While the job slot matrices show the resource profiles of the queuing jobs. For example, the red dots indicate a job is scheduled at the next timestep for machine 2, it requests 2 units of CPU resource and 1 unit of memory resource and lasts 3 time units.

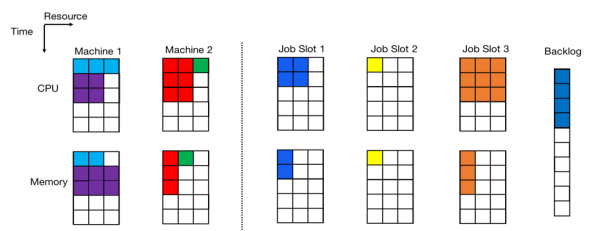

Figure 1. State Representation

**Action Space.** Different from the large size of the state space, with a fixed number of machines and queue length, the action space could be well designed to be small enough. We define the action $i*q + j$ as "assign job $j$ in queue to machine $i$", where $q$ is the length of the job queue. A "void" action indicates the agent does not schedule further jobs at the current timestep. Hence, the size of action space is $m*q + 1$, where $m$ is the number of machines in the cluster. At each timestep, the scheduler could take multiple actions until it chooses the void action or an invalid action (e.g. attempting to schedule a job that does not "fit"). With each valid action, one job is scheduled at the first possible timestep in the machine and the system state changes. After a void or invalid action is chosen, time proceeds so that new jobs enter the queue and jobs in the machines get processed.

**Rewards.** In our case, rewards are designed so that the system learns to minimize the average slowdown. Therefore, we relate it with the inverse of time length of each job request. Specially, we define the reward for each action as below:

$$Reward = -\left(\sum_l \frac{\alpha_l}{\sum_{i_l} t_{i_l}} + \frac{\beta}{\sum_j t_j} + \frac{\gamma}{\sum_k t_k}\right), \quad (1)$$

where $l \in$ all machines in the cluster, $i_l \in$ all scheduled jobs for machines $l$, $j \in$ all jobs in the queue, and $k \in$ all jobs in the backlog. Notice that setting the discount factor = 1, and $\alpha_l, \beta, \gamma$ = 1, the cumulative reward over time coincides with (negative) the sum of job slowdowns, hence minimizing the average job slowdown. In our model, wanting jobs to be finished as soon as possible, we give greater penalty for jobs in the queue by assigning a slightly bigger value for $\beta$. We also design different $\alpha_l$ to denote the different transmission speed from job queue to different machines.

**Neural Network Settings.** We constructed a multi-layer convolutional neural network (CNN) to extract features from the state matrix, as CNN is proven to do well in extracting high-level features from images. Table I shows the architecture of our neural network.

**Training Algorithm.** We adopt the same training algorithm as DeepRM, using policy gradient methods.

### III. EVALUATION

We use the same workload as DeepRM [3]. For the extended model of scheduling for multiple machines, we assume having two machines in the cluster. In Fig. 2, we showed the training curves of our model for single machine task scheduling in comparison with the original DeepRM scheduler. Fig. 2(a) depicts that by reshaping the state space, our model (DeepRM-reshape) outperforms DeepRM by reducing the average job slowdown by up to 8.57%. Fig. 2(b) demonstrates that using a CNN as input layer, we end up with a faster convergence rate during training as well as a decrease in average job slowdown by 4.78%. Fig. 2(c) reports that our modified reward function (setting $\alpha_l$=1, $\beta$=2, $\gamma$=1) does not improve job slowdown, but generates a slightly faster convergence rate. Fig. 3 reports the discounted total reward and average job slowdown of our extended model after 100 training iterations. Not surprisingly, our model outperforms heuristic methods such as Shortest Job First (SJF) and Packer.

TABLE I. NEURAL NETWORK ARCHITECTURE

| Layer | Convolutional (Input) | Fully-connected (Output) |
|---|---|---|
| Input size | 124*20 = 2480 | 122*18 = 2196 |
| Fiter size | 2×2 | ―― |
| Stride | (1,1) | ―― |
| #Filters | 16 | 6 |
| Activation | Relu | Softmax |
| Output size | 122*18 = 2196 | 6(queue length+1) |

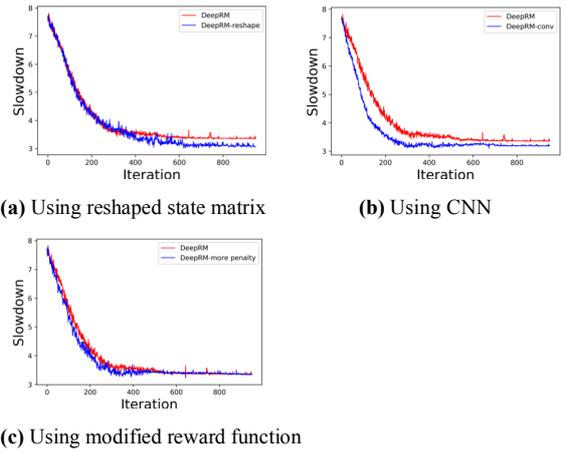

(a) Using reshaped state matrix  (b) Using CNN

(c) Using modified reward function

Figure 2. Comparison of learning curves of average slowdown generated by DeepRM and three different improvements based on it

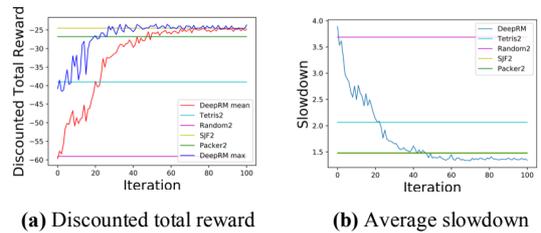

(a) Discounted total reward  (b) Average slowdown

Figure 3. Learning curve showing that the scheduling agent increases the discounted total reward and lowers the average slowdown after epochs of training.

### IV. DISCUSSION AND FUTURE WORK

In this paper, we explored a reinforcement learning method to solve the cluster scheduling problem. In the future, we want to take into consideration factors such as the dependency between jobs, the locality issue of machines and having multiple queues to mimic multiple users.